\documentclass[aps,prd,superscriptaddress,showpacs,preprint]{revtex4}
\usepackage{graphicx, bm}
\usepackage[usenames]{color}
\usepackage{float}

\begin{document}

\draft
\title{The $\mu^+\mu^-$ collider to sensitivity estimates on the magnetic and electric dipole moments of the tau-lepton}

\author{M. K\"{o}ksal\footnote{mkoksal@cumhuriyet.edu.tr}}
\affiliation{\small Deparment of Optical Engineering, Sivas Cumhuriyet University, 58140, Sivas, Turkey.\\}

\author{A. A. Billur\footnote{abillur@cumhuriyet.edu.tr}}
\affiliation{\small Deparment of Physics, Sivas Cumhuriyet University, 58140, Sivas, Turkey.\\}

\author{ A. Guti\'errez-Rodr\'{\i}guez\footnote{alexgu@fisica.uaz.edu.mx}}
\affiliation{\small Facultad de F\'{\i}sica, Universidad Aut\'onoma de Zacatecas\\
         Apartado Postal C-580, 98060 Zacatecas, M\'exico.\\}

\author{ M. A. Hern\'andez-Ru\'{\i}z\footnote{mahernan@uaz.edu.mx}}
\affiliation{\small Unidad Acad\'emica de Ciencias Qu\'{\i}micas, Universidad Aut\'onoma de Zacatecas\\
         Apartado Postal C-585, 98060 Zacatecas, M\'exico.\\}

\date{\today}

\begin{abstract}

Using the effective Lagrangian formalism, the anomalous magnetic and electric dipole moments of the tau-lepton in the
$\mu^{+}\mu^{-} \rightarrow \mu^{+}\gamma^{*} \gamma^{*} \mu^{-}\rightarrow \mu^{+}\tau \bar{\tau}\mu^{-}$ process at the future
muon colliders at the $\sqrt{s}=1.5, 3$ and $6$ TeV are investigated. In addition, the bounds at the $95\%$ confidence level on
the dipole moments of the tau-lepton using different integrated luminosities are estimated. It is shown that the $\mu^{+}\mu^{-} \rightarrow \mu^{+}\gamma^{*} \gamma^{*} \mu^{-}\rightarrow \mu^{+}\tau \bar{\tau}\mu^{-}$ process leads to a remarkable improvement in the
existing experimental bounds on the anomalous magnetic and electric dipole moments of the tau-lepton.

\end{abstract}


\vspace{5mm}

\maketitle

\section{Introduction}

A high priority on the physic program for the current and future of High Energy Physics (HEP) is the quest for physics Beyond
the Standard Model (BSM). With this motivation a $\mu^+\mu^-$ collider at the CERN, is one of the potential candidates  for a
future energy frontier colliding machine. The original idea about the possibility of muon colliders was proposed by G.I.
Budker \cite{Budker}, Skrinsky and Parkhomchuk \cite{Parkhomchuk} and Neuffer \cite{Neuffer}. More recently, a collaboration of
different members, has been formed to coordinate studies on specific designs \cite{Charles,Eichten,Palmer}. The design of this
collider demonstrates a novel high-energy and high-luminosity collider type, which will permit exploration of HEP at energy
frontiers beyond the reach of currently existing and proposed electron-positron colliders. In addition, the proposed 1.5, 3, 6 TeV
center-of-mass energies $\mu^+\mu^-$ collider in the CERN provides outstanding discovery potential and can complement the physics
program of the Large Hadron Collider (LHC).

The reason as well as the advantage for interest in muon colliders is that they are fundamental leptons with a mass that is a factor
of 207 greater than the mass of the electron or positron. As for an electron, the full center-of-mass energy is available in an interaction. But
because of the large mass, there is essentially no synchrotron radiation from the muon in comparison to electrons and positrons.
Consequently, the machine can be circular and much smaller than the current design of linear electron-positron colliders, and the hope
is that the sum of development and construction costs will not be so high as to make the realization unaffordable.

A muon collider will accelerate two muon beams in opposite directions around an underground ring 6.3 km of circumference.
Beams will collide head-on and scientists will study what results from the collision to search for dark matter, dark energy,
the matter-antimatter asymmetry, supersymmetric particles, signs of extra dimensions and other subatomic phenomena. Furthermore,
a muon collider has the characteristic that it focuses on a region of energy to discover the physical phenomena that the LHC can
not reveal on its own. A muon collider would provide a clear and unobstructed view of the subatomic world. In addition, the beauty
of a muon collider is that the collision events are clean.

Starting from the feasibility of a muon collider to study new physics, we study the anomalous Magnetic Moment ($\tau$MM) and Electric
Dipole Moment ($\tau$EDM) of the tau-lepton in the $\mu^{+}\mu^{-} \rightarrow \mu^{+}\gamma^{*} \gamma^{*} \mu^{-}\rightarrow
\mu^{+}\tau \bar{\tau}\mu^{-}$ process at the future muon collider at the $\sqrt{s}=1.5, 3$ and $6$ TeV. In addition, the bounds at the
$95\%$ Confidence Level (C.L.) on the dipole moments of the tau-lepton using different integrated luminosities ${\cal L}=10, 20, 50, 100, 200,
300, 400, 500, 710 \hspace{0.8mm}fb^{-1}$ and systematic uncertainties of $\delta_{sys}=0\%, 3\%, 5\%$ are estimated.

In our study, the quasi-real photons in $\gamma^{*}\gamma^{*}$ collisions can be examined
by Equivalent Photon Approximation (EPA) \cite{Budnev,Baur,Piotrzkowski}, that is to say, using the Weizsacker-Williams
approximation (WWA). In EPA, photons emitted from incoming leptons which have very low virtuality are scattered
at very small angles from the beam pipe and because the emitted quasi-real photons have a low $Q^{2}$ virtuality,
these are almost real. These processes have been observed experimentally at the LEP, Tevatron and LHC \cite{Abulencia,Aaltonen1,Aaltonen2,Chatrchyan1,Chatrchyan2,Abazov,Chatrchyan3}. In  particular, the most stringent
experimental limit on the anomalous $\tau$MM and $\tau$EDM is obtained through the process
$e^{+}e^{-} \rightarrow e^{+}\gamma^{*} \gamma^{*} e^{-} \rightarrow e^{+} \tau \bar{\tau} e^{-}$ by using
multiperipheral collision at the LEP \cite{DELPHI}.

The study for the dipole moments is a very active field with ongoing experiments to measure the dipole moments of a variety
of physical systems such as Atoms, Molecules, Nuclei and Particles, see e.g. Refs. \cite{Engel,Yamanaka,Chupp} for recent reviews.
In addition, the dipole moments of particles can be probed by analyzing decay and collision processes. This has been done for the
tau-lepton in processes such as $e^+e^- \to e^+e^-\tau^+\tau^-$ by DELPHI Collaboration \cite{DELPHI} and $e^+e^- \to \tau^+\tau^-$
by BELLE Collaboration \cite{BELLE}, respectively, obtaining the followings bounds:

\begin{equation}
\mbox{DELPHI}:
\begin{array}{ll}
-0.052 < a_\tau < 0.013, \hspace{3mm}  & \mbox{$95\%$ C.L.}, \\
-3.7 < d_\tau(10^{-16}e cm) < 3.7, \hspace{3mm}  & \mbox{$95\%$ C.L.},
\end{array}
\end{equation}

\noindent and

\begin{equation}
\mbox{BELLE}:
\begin{array}{ll}
-2.2 < Re(d_\tau(10^{-17}e cm)) < 4.5, \hspace{3mm}  & \mbox{$95\%$ C.L.}, \\
-2.5 < Im(d_\tau(10^{-17}e cm)) < 0.8, \hspace{3mm}  & \mbox{$95\%$ C.L.}.
\end{array}
\end{equation}

A summary of experimental and theoretical bounds on the dipole moments of the $\tau$-lepton are given in Table I of Ref. \cite{Koksal1}.
See Refs. \cite{Bernreuther0,Iltan1,Dutta,Iltan2,Iltan3,Iltan4,Gutierrez1,Gutierrez2,Gutierrez3,Gutierrez4,Ozguven,Billur,Sampayo,
Passera1,Eidelman1,Koksal3,Arroyo1,Arroyo2,Xin,Pich,Atag1,Lucas,Passera2,Passera3,Bernabeu,Bernreuther} for another
bounds on the $\tau$MM and the $\tau$EDM in different context.

The paper is organized in the following way. In Section II, are given the gauge-invariant operators of dimension six. In Section III, we study the
total cross-section and the dipole moments of the tau-lepton through the process $\mu^{+}\mu^{-} \rightarrow \mu^{+}\gamma^{*} \gamma^{*} \mu^{-}\rightarrow \mu^{+}\tau \bar{\tau}\mu^{-}$ at the $\gamma^*\gamma^*$ collision mode. Finally, we present our conclusions in Section IV.

\vspace{5mm}

\section{Electromagnetic current and operators of dimension six}

\subsection{$\tau^+\tau^-\gamma$ vertex form factors}

The proposed high-energy and high-luminosity $\mu^+\mu^-$ collider offers new opportunities for the improved determination
of the fundamental physical parameters of standard heavy leptons. In this sense, compared to the electron or the muon case,
the electromagnetic properties of the $\tau$-lepton are largely unexplored. On this topic, a convenient way of
studying its electromagnetic properties on a model-independent way is through the effective
tau-photon interaction vertex which is described by four independent form factors.
The possible electromagnetic properties of the $\tau$-lepton are summarized in the most general expression
consistent with Lorentz and electromagnetic gauge invariance for the $\tau^+\tau^-\gamma$
vertex between on-shell tau-lepton and the photon \cite{Passera1,Grifols,Escribano,Giunti,Giunti1} as follows

\begin{equation}
\Gamma^{\alpha}_\tau=eF_{1}(q^{2})\gamma^{\alpha}+\frac{ie}{2m_\tau}F_{2}(q^{2})\sigma^{\alpha\mu}q_{\mu}
+ \frac{e}{2m_\tau}F_3(q^2)\sigma^{\alpha\mu}q_\mu\gamma_5 +eF_4(q^2)\gamma_5(\gamma^\alpha - \frac{2q^\alpha m_\tau}{q^2}).
\end{equation}

\noindent The quantities $e$ and $m_\tau$ are the charge of the electron and the mass of the $\tau$-lepton, respectively.
$\sigma^{\alpha\mu}=\frac{i}{2}[\gamma^\alpha, \gamma^{\mu}]$ and $q=p'-p$ is the four-momentum of the photon.
The form factors $F_{1,2,3,4}(q^2)$ have the following interpretations for $q^2=0$:

\begin{eqnarray}
Q_\tau &=& F_1(0), \hspace{5mm} \mbox{Electric charge}, \\
a_\tau &=& F_2(0),  \hspace{5mm} \mbox{$\tau$MM}, \\
d_\tau &=&\frac{e}{2m_\tau} F_3(0)   , \hspace{5mm} \mbox{$\tau$EDM}.
\end{eqnarray}

\noindent $F_4(q^2)$ is the Anapole form factor.

\subsection{Gauge-invariant operators of dimension six}

In theoretical, experimental and phenomenological searches most of the tau-lepton electromagnetic vertices search involve off-shell tau-leptons.
In our study, one of the tau-leptons is off-shell and measured quantity is not directly $a_\tau$ and $d_\tau$. For this reason deviations
of the tau-lepton dipole moments from the SM values are examined in a model independent way using the effective Lagrangian formalism.
This formalism is defined by high-dimensional operators which lead to anomalous $\tau^+ \tau^- \gamma$ coupling. For our study, we
apply the dimension-six effective operators that contribute to the $\tau$MM and $\tau$EDM \cite{Buchmuller,1,eff1,eff3}:

\begin{eqnarray}
L_{eff}=\frac{1}{\Lambda^{2}} \Bigl[C_{LW}^{33} Q_{LW}^{33}+C_{LB}^{33} Q_{LB}^{33} + \mbox{h.c}\Bigr],
\end{eqnarray}

\noindent where

\begin{eqnarray}
Q_{LW}^{33}=\bigl(\bar{\ell_{\tau}}\sigma^{\mu\nu}\tau_{R}\bigr)\sigma^{I}\varphi W_{\mu\nu}^{I},
\end{eqnarray}

\begin{eqnarray}
Q_{LB}^{33}=\bigl(\bar{\ell_{\tau}}\sigma^{\mu\nu}\tau_{R}\bigr)\varphi B_{\mu\nu},
\end{eqnarray}

\noindent in which, $B_{\mu\nu} $ is the $U(1)_Y$ gauge field strength tensors and $W_{\mu\nu}^{I}$ is the $SU(2)_L$
gauge field strength tensors, respectively, while $\varphi$ and $\ell_{\tau}$ are the Higgs and the left-handed $SU(2)_L$
doublets which contain $\tau$, and $\sigma^{I}$ are the Pauli matrices.

The corresponding CP even $\kappa$ and CP odd $\tilde{\kappa}$ observables are obtained with
the electroweak symmetry breaking from the effective Lagrangian given by Eq. (7):

\begin{eqnarray}
\kappa=\frac{2 m_{\tau}}{e} \frac{\sqrt{2}\upsilon}{\Lambda^{2}} Re\Bigl[\cos\theta _{W} C_{LB}^{33}- \sin\theta _{W} C_{LW}^{33}\Bigr],
\end{eqnarray}

\begin{eqnarray}
\tilde{\kappa}=\frac{2 m_{\tau}}{e} \frac{\sqrt{2}\upsilon}{\Lambda^{2}} Im\Bigl[\cos\theta _{W} C_{LB}^{33}- \sin\theta _{W} C_{LW}^{33}\Bigr],
\end{eqnarray}

\noindent where $\upsilon=246$ GeV is the breaking scale of the electroweak symmetry, $\Lambda$ is the new physics scale and
$\sin\theta _{W}$ is the sin of the weak mixing angle.

These observables are related to contribution of the anomalous $\tau$MM and $\tau$EDM through the following relations:

\begin{eqnarray}
\kappa&=&\tilde{a}_{\tau},  \\
\tilde{\kappa}&=&\frac{2m_{\tau}}{e}\tilde{d}_{\tau}.
\end{eqnarray}

\vspace{5mm}

\section{The cross-section of the process $\mu^{+}\mu^{-} \rightarrow \mu^{+}\gamma^{*} \gamma^{*} \mu^{-}\rightarrow \mu^{+}\tau \bar\tau\mu^{-}$}

In order to study the opportunities of the muon collider as an option to sensitivity estimates on the $\tau$MM and $\tau$EDM
in detail, we focus here only on the cross-section of the process $\mu^{+}\mu^{-} \rightarrow \mu^{+}\gamma^{*} \gamma^{*} \mu^{-}\rightarrow \mu^{+}\tau \bar{\tau}\mu^{-}$. The Feynman diagrams at the tree level are given in Fig. 1.

\subsection{$\gamma^{*} \gamma^{*}  \to \tau^+\tau^-$ cross-section}

The corresponding matrix elements for the subprocess $\gamma^*\gamma^* \to \tau^+\tau^-$ in terms of the Mandelstam invariants
$\hat s$, $\hat t$, $\hat u$ and from the anomalous parameters $\kappa$ and $\tilde{\kappa}$ are

\begin{eqnarray}
|M_{1}|^{2}&=&\frac{16\pi^{2}Q_{\tau}^2\alpha^{2}_e}{2m_{\tau}^{4}(\hat{t}-m_{\tau}^{2})^{2}}\biggl[48 \kappa (m_{\tau}^{2}-\hat{t})
(m_{\tau}^{2}+\hat{s}-\hat{t})m_{\tau}^{4}-16(3m_{\tau}^{4}-m_{\tau}^{2}\hat{s}+\hat{t}(\hat{s}+\hat{t})) m_{\tau}^{4}\nonumber\\
&+&2(m_{\tau}^{2}-\hat{t})( \kappa^{2}( 17m_{\tau}^{4}+(22\hat{s}-26\hat{t})m_{\tau}^{2} +\hat{t}(9\hat{t}-4\hat{s}))  \nonumber\\
&+&\tilde{\kappa}^{2}(17m_{\tau}^{2}+4\hat{s}-9\hat{t})(m_{\tau}^{2}-\hat{t}))m_{\tau}^{2}+12\kappa(\kappa^{2}
+\tilde{\kappa}^{2})\hat{s}(m_{\tau}^{3}-m_{\tau}\hat{t})^{2}\nonumber\\
&-&(\kappa^{2}+ \tilde{\kappa}^{2})^{2}(m_{\tau}^{2}-\hat{t})^{3}(m_{\tau}^{2}-\hat{s}-\hat{t})\biggr],
\end{eqnarray}

\begin{eqnarray}
|M_{2}|^{2}&=&\frac{-16\pi^{2}Q_{\tau}^2\alpha^{2}_e}{2m_{\tau}^{4}(\hat{u}-m_{\tau}^{2})^{2}}\biggl[48 \kappa (m_{\tau}^{4}+(\hat{s}-2\hat{t})m_{\tau}^{2}+\hat{t}(\hat{s}+\hat{t}))m_{\tau}^{4}\nonumber\\
&+&16(7m_{\tau}^{4}-(3\hat{s}+4\hat{t})m_{\tau}^{2}+\hat{t}(\hat{s}+\hat{t})) m_{\tau}^{4}\nonumber\\
&+&2(m_{\tau}^{2}-\hat{t})( \kappa^{2}(m_{\tau}^{4}+(17\hat{s}-10\hat{t})m_{\tau}^{2}+9\hat{t}(\hat{s}+\hat{t})) \nonumber\\
&+& \tilde{\kappa}^{2}(m_{\tau}^{2}-9\hat{t})(m_{\tau}^{2}-\hat{t}-\hat{s}))m_{\tau}^{2}\nonumber\\
&+&( \kappa^{2}+ \tilde{\kappa}^{2})^{2}(m_{\tau}^{2}-\hat{t})^{3}(m_{\tau}^{2}-\hat{s}-\hat{t})\biggr],
\end{eqnarray}

\begin{eqnarray}
M_{1}^{\dag}M_{2}+M_{2}^{\dag}M_{1}&=&\frac{16\pi^{2}Q_{\tau}^2\alpha^{2}_e}{m_{\tau}^{2}(\hat{t}-m_{\tau}^{2})(\hat{u}-m_{\tau}^{2})} \nonumber \\
&\times &\biggl[-16(4m_{\tau}^{6}-m_{\tau}^{4}\hat{s})+8 \kappa m_{\tau}^{2}(6m_{\tau}^{4}-6m_{\tau}^{2}(\hat{s}+2\hat{t})-\hat{s})^{2} \nonumber \\ &+&6\hat{t})^{2}+6\hat{s}\hat{t})+( \kappa^{2}(16m_{\tau}^{6}-m_{\tau}^{4}(15\hat{s}+32\hat{t})+m_{\tau}^{2}(15\hat{s})^{2} \nonumber \\
&+&14\hat{t}\hat{s}+16\hat{t})^{2})+\hat{s}\hat{t}(\hat{s}+\hat{t}))+ \tilde{\kappa}^{2}(16m_{\tau}^{6}-m_{\tau}^{4}(15\hat{s}+32\hat{t})  \nonumber\\
&+&m_{\tau}^{2}(5\hat{s})^{2}+14\hat{t}\hat{s}+16\hat{t})^{2})+\hat{s}\hat{t}(\hat{s}+\hat{t})))-4 \kappa \hat{s}( \kappa^{2}+ \tilde{\kappa}^{2})\nonumber\\
&\times& (m_{\tau}^{4}+m_{\tau}^{2}(\hat{s}-2\hat{t})+\hat{t}(\hat{s}+\hat{t}))  \nonumber\\
&&-2\hat{s}(\kappa^{2}+ \tilde{\kappa}^{2})^{2}(m_{\tau}^{4}-2\hat{t}m_{\tau}^{2}+\hat{t}(\hat{s}+\hat{t}))\biggr].
\end{eqnarray}

Here, the Mandelstam variables are $\hat s=(p_1 + p_2)^2=(p_3 + p_4)^2$, $\hat t=(p_1 - p_3)^2=(p_4 - p_2)^2$, $\hat u=(p_3 - p_2)^2=(p_1 - p_4)^2$,
while $p_{1}$ and $p_{2}$ are the four-momenta of the incoming photons, $p_{3}$ and $p_{4}$ are the momenta of the outgoing tau-lepton,
$Q_{\tau}$ is the tau-lepton charge and $\alpha_e$ is the fine-structure constant.

WWA is another possibility for tau pair production, and the quasi-real photons emitted from both lepton beams collide with each other
and produce the subprocess $\gamma^{*} \gamma^{*} \rightarrow \tau^+ \tau^-$. In WWA, the photon spectrum is given by

\begin{eqnarray}
f_{\gamma^{*}}(x)=\frac{\alpha}{\pi E_{\mu}}\lbrace [\frac{1-x+x^2/2}{x}]log(\frac{Q^2_{max}}{Q^2_{min}})-\frac{m_{\mu}^2x}{Q^2_{min}}(1-\frac{Q^2_{min}}{Q^2_{max}})-\frac{1}{x}
[1-\frac{x}{2}]^2log(\frac{x^2 E_{\mu}^2+Q^2_{max}}{x^2 E_{\mu}^2+Q^2_{min}})\rbrace,
\end{eqnarray}

\noindent where $x=E_{\gamma}/E_{\mu}$ and $Q^2_{max}$ is maximum virtuality of the photon. In this work, we have taken into account
the maximum virtuality of the photon as $Q^2_{max}=2, 16, 64 \hspace{0.8mm}GeV^2$. The minimum value of the $Q^2_{min}$ is given by

\begin{eqnarray}
Q^2_{min}=\frac{m_{\mu}^2x^2}{1-x}.
\end{eqnarray}

The reaction $\gamma^{*} \gamma^{*} \rightarrow \tau^+ \tau^-$ participates as a subprocess in the main process $\mu^{+}\mu^{-} \rightarrow
\mu^{+}\gamma^{*} \gamma^{*} \mu^{-}\rightarrow \mu^{+}\tau \bar{\tau}\mu^{-}$, and the total cross-section is given by

\begin{eqnarray}
\sigma=\int f_{\gamma^*}(x)f_{\gamma^*}(x)d\hat{\sigma}dE_{1}dE_{2}.
\end{eqnarray}

We presented results for the dependence of the total cross-section of the process $\mu^{+}\mu^{-} \rightarrow
\mu^{+}\gamma^{*} \gamma^{*} \mu^{-}\rightarrow \mu^{+}\tau \bar{\tau}\mu^{-}$ on $\kappa (\tilde\kappa)$. We consider
the following cases with $Q^2_{max}=2, 16, 64 \hspace{0.8mm}GeV^2$:\\

\newpage

$\bullet$ For $\sqrt{s}=1500\hspace{0.8mm} GeV$.

\begin{eqnarray}
\sigma(\kappa)&=&\Bigl[(5.64\times10^4; 1.02\times10^5; 1.42\times10^5)\kappa^4 + (4.43\times10^2; 6.93\times10^2; 8.83\times10^2)\kappa^3  \nonumber\\
&+& (4.45\times10^2; 7.00\times10^2; 8.87\times10^2)\kappa^2 +(1.14; 1.70; 2.13)\kappa \nonumber\\
&+&(0.35; 0.52; 0.65)\Bigr] (pb)   \\
\sigma(\tilde{\kappa})&=&\Bigl[(5.64\times10^4; 1.02\times10^5; 1.42\times10^5)\tilde{\kappa}^4 + (4.45\times10^2; 7.00\times10^2; 8.87\times10^2)\tilde{\kappa}^2   \nonumber\\
&+& (0.35; 0.52; 0.65) \Bigr] (pb).
\end{eqnarray}

$\bullet$ For $\sqrt{s}=3000\hspace{0.8mm} GeV$.

\begin{eqnarray}
\sigma(\kappa)&=&\Bigl[(2.30\times10^5; 4.18\times10^5; 5.75\times10^5)\kappa^4 + (7.54\times10^2; 1.11\times10^2; 1.40\times10^2)\kappa^3  \nonumber\\
&+& (7.59\times10^2; 1.13\times10^3; 1.43\times10^3)\kappa^2 +(1.65; 2.34; 2.86)\kappa \nonumber\\
&+&(0.51; 0.72; 0.88)\Bigr] (pb)   \\
\sigma(\tilde{\kappa})&=&\Bigl[(2.30\times10^5; 4.18\times10^5; 5.75\times10^5)\tilde{\kappa}^4 + (7.59\times10^3; 1.13\times10^3; 1.43\times10^3)\tilde{\kappa}^2   \nonumber\\
&+& (0.51; 0.72; 0.88) \Bigr] (pb).
\end{eqnarray}

$\bullet$ For $\sqrt{s}=6000\hspace{0.8mm} GeV$.

\begin{eqnarray}
\sigma(\kappa)&=&\Bigl[(9.23\times10^5; 1.68\times10^6; 2.31\times10^6)\kappa^4 + (1.06\times10^3; 1.68\times10^3; 2.07\times10^3)\kappa^3  \nonumber\\
&+& (1.17\times10^3; 1.69\times10^3; 2.09\times10^3)\kappa^2 +(2.26; 3.00; 3.60)\kappa \nonumber\\
&+&(0.68; 0.93; 1.11)\Bigr] (pb)   \\
\sigma(\tilde{\kappa})&=&\Bigl[(9.23\times10^5; 1.68\times10^6; 2.31\times10^6)\tilde{\kappa}^4 + (1.17\times10^3; 1.69\times10^3; 2.09\times10^3)\tilde{\kappa}^2   \nonumber\\
&+& (0.68; 0.93; 1.11) \Bigr] (pb).
\end{eqnarray}

These formulas have been obtained with the help of the package CALCHEP \cite{Belyaev}, which can computate the Feynman
diagrams, integrate over multiparticle phase space and event simulation. Furthermore, we apply the following acceptance
cuts for $\tau^+\tau^-$ signal at the muon collider:

\newpage

\begin{eqnarray}
\begin{array}{l}
p^{\tau, \bar\tau}_t > 20 \hspace{0.9mm}GeV, \hspace{0.8mm}  {\mbox{(transverse momentum of the final state particles)}},\\
|\eta^{\tau, \bar\tau}| < 2.5, \hspace{0.9mm}  {\mbox{(pseudorapidity reduces the contamination from other particles}} \\
\hspace*{2.3cm}                                                {\mbox{misidentified as tau)}}, \\
\Delta R(\tau, \bar\tau ) > 0.4, \hspace{0.9mm}  {\mbox{(separation of the final state particles)}}, \\
\end{array}
\end{eqnarray}

\noindent of course, is fundamental that we apply these cuts to reduce the background and to optimize the signal sensitivity.

From Eqs. (20)-(25), the dependent terms on $\kappa (\tilde{\kappa})$ are purely anomalous, and the independent
term of $\kappa (\tilde{\kappa})$ give the cross-section of the SM.

\subsection{Sensitivity on the ${\tilde a}_\tau$ and ${\tilde d}_\tau$ through $\mu^{+}\mu^{-} \rightarrow \mu^{+}\gamma^{*}
\gamma^{*} \mu^{-}\rightarrow \mu^{+}\tau \bar{\tau}\mu^{-}$ at the muon collider}

A muon collider is an ideal discovery machine in the multi-TeV energy range. In this subsection, we assess the capabilities of future
muon collider to test the existence of the $\tau$MM and $\tau$EDM by means of the process $\mu^{+}\mu^{-} \rightarrow \mu^{+}\gamma^{*}
\gamma^{*} \mu^{-}\rightarrow \mu^{+}\tau \bar{\tau}\mu^{-}$ at the mode $\gamma^{*} \gamma^{*} \rightarrow \tau \bar{\tau}$.
Specifically, we assume energies from 1.5 to 6 TeV and integrated luminosities of at least $10-710\hspace{0.8mm}fb^{-1}$.
The primary motivation for the $\sqrt{s}=1.5, 3, 6\hspace{0.8mm}TeV$ center of-mass energies, luminosities
${\cal L}=10, 20, 50, 100, 300, 500, 710\hspace{0.8mm}fb^{-1}$ of such a collider, as well as of the virtuality of the photon
$Q^2_{max}=2, 16, 64\hspace{0.8mm}GeV^2$ is to optimize the expected signal cross-section and the sensitivity on
${\tilde a}_\tau$ and ${\tilde d}_\tau$.

The total cross-section production as a function of $\kappa$(${\tilde\kappa}$) has been computed in the previous section (see Eqs. (20)-(25)),
and is displayed in Figs. 2-7. The graphics are for $Q^2_{max}=2, 16, 64\hspace{0.8mm}GeV^2$ with $\sqrt{s}=1.5, 3, 6\hspace{0.8mm}TeV$.
These figures clearly show a strong dependence with respect to $\kappa$(${\tilde\kappa}$), the virtuality of the photon $Q^2_{max}$,
as well as with the center-of-mass energy $\sqrt{s}$. In Figs. 8-13, the graphics are for $\sqrt{s}=1.5, 3, 6\hspace{0.8mm}TeV$ with
$Q^2_{max}=2, 16, 64 \hspace{0.8mm}GeV^2$, respectively. These figures also show a clear and strong dependence with respect to
$\kappa$(${\tilde\kappa}$), $\sqrt{s}$ and $Q^2_{max}$. The total cross-section increase of the order of $\sigma=20\hspace{0.8mm}pb$
at the upper and lower limit of $\kappa$($\tilde{\kappa}$) and tends to the value of the SM when $\kappa$(${\tilde\kappa}$) tends to
zero, as indicated by Eqs. (20)-(25).

To estimate the sensitivity on the parameters ${\tilde a}_\tau$ and ${\tilde d}_\tau$ we consider the acceptance cuts given in Eq. (26),
take into account the systematic uncertainties $\delta_{sys} = 0, 3, 5 \%$ and we adopt the statistical method for the $\chi^2$
defined as \cite{Koksal1,Koksal2,Gutierrez13,Koksal,Ozguven,Sahin1,Billur1,Billur2}

\begin{equation}
\chi^2=\Biggl(\frac{\sigma_{SM}-\sigma_{BSM}(\sqrt{s}, Q^2_{max}, \tilde a_\tau, \tilde d_\tau)}{\sigma_{SM}\sqrt{(\delta_{st})^2
+(\delta_{sys})^2}}\Biggr)^2,
\end{equation}

\noindent with $\sigma_{BSM}(\sqrt{s}, Q^2_{max}, \tilde a_\tau, \tilde d_\tau)$ the total cross-section incorporating
contributions from the SM and new physics, $\delta_{st}=\frac{1}{\sqrt{N_{SM}}}$ is the statistical error
and $\delta_{sys}$ is the systematic error. The number of events is given by $N_{SM}={\cal L}_{int}\times \sigma_{SM}$,
where ${\cal L}_{int}$ is the integrated luminosity of the $\mu^+\mu^-$ collider.

We now discuss the reach at different center-of-mass energies in the sensitivity estimates determination on the $\tau$MM and $\tau$EDM.
As we will show below, the sensitivity of the electromagnetic properties of the tau-lepton, and in
particular its magnetic and electric dipole moments, may be measured competitively in these facilities, using the process $\mu^{+}\mu^{-}
\rightarrow \mu^{+}\gamma^{*} \gamma^{*} \mu^{-} \rightarrow \mu^{+}\tau \bar{\tau}\mu^{-}$. The sensitivities from the ${\tilde a}_\tau$
and ${\tilde d}_\tau$ turn out to be very strong at $\mu^+\mu^-$ collider. For this reason we performing a detailed study and we present
Tables that illustrate the sensitivity on ${\tilde a}_\tau$ and ${\tilde d}_\tau$ for different virtuality of the photon $Q^2_{max}$,
center-of-mass energies $\sqrt{s}$, luminosities ${\cal L}$, uncertainties systematic $\delta_{sys}$ and at $95\%$ C.L. Our results are
presented in Tables I-III. Our most significant results on ${\tilde a}_\tau$ and ${\tilde d}_\tau$ are the following for $\sqrt{s}=6\hspace{0.8mm}TeV$,
${\cal L}=710\hspace{0.8mm}fb^{-1}$, $\delta_{sys}=0$ and $Q^2_{max}=2, 16, 64\hspace{0.8mm}GeV^2$,
${\tilde a}_\tau=(-0.00278, 0.00086); (-0.00253, 0.00076); (-0.00242, 0.00071)$, respectively, at $95\%$ C.L.. In the case of the electric
dipole moment, our most important results are $|{\tilde d}_\tau|= (0.863,  0.777,  0.731)\times 10^{-17}$ at $95\%$ C.L..
From these results it is seen that the sensitivity improves for large values of $Q^2_{max}$. In addition, notice that these results give
an improvement of 1-2 orders of magnitude with respect to the results given in Eqs. (1) and (2) obtained by the DELPHI and BELLE Collaborations.

Furthermore, in order to following study the opportunities of the muon collider in detail, we focus now in the bounds contours on the
$(\kappa, \tilde\kappa)$ plane depending on integrated luminosity and for $\sqrt{s}=1.5, 3, 6 \hspace{0.8mm}TeV$ and
$Q^2_{max}=2\hspace{0.8mm}GeV^2$ in Figs. 17-19. The sensitivity reach of a muon collider is indicated by a blue, yellow and green solid
line in each plot. These results show that anomalous couplings $\kappa$ and $\tilde\kappa$ can be probed with very good sensitivity in a
muon collider.

\begin{table}[!ht]
\caption{Model-independent sensitivity estimates for the $\tilde{a}_\tau$ magnetic moment and the $\tilde{d}_\tau$ electric dipole moment
for $Q^2_{max}=2, 16, 64\hspace{0.8mm}GeV^2$, $\sqrt{s}=1.5\hspace{0.8mm}TeV$ and ${\cal L}=10, 20, 50, 100, 110\hspace{0.8mm}fb^{-1}$ at $95\%$ C.L.,
through the process $\mu^{+}\mu^{-} \rightarrow \mu^{+}\gamma^{*} \gamma^{*} \mu^{-} \rightarrow \mu^{+}\tau \bar{\tau}\mu^{-}$.}
\begin{center}
 \begin{tabular}{ccccc}
\hline\hline
\multicolumn{4}{c}{ $\sqrt{s}=1.5\hspace{0.8mm}TeV$, \hspace{5mm}  $95\%$ C.L.}\\
 \hline
 \cline{1-4}  ${\cal L}\hspace{0.8mm}(fb^{-1})$  &  $\delta_{sys}$ &    $\tilde{a}_\tau$   &  $|\tilde{d}_\tau(e cm)|$ \\
\hline
10  &  $0\%$   & [(-0.00763; -0.00684; -0.00646), (0.00502; 0.00438; 0.00405)]   &       $(3.445; 3.032; 2.845)\times 10^{-17}$   \\
10  &  $3\%$   & [(-0.00916; -0.00862; -0.00839), (0.00654; 0.00616; 0.00598)]   &       $(4.312; 4.039; 3.941)\times 10^{-17}$   \\
10  &  $5\%$   & [(-0.01066; -0.01020; -0.01000), (0.00804; 0.00773; 0.00758)]   &       $(5.155; 4.920; 4.842)\times 10^{-17}$   \\
\hline
20  & $0\%$    & [(-0.00667; -0.00599; -0.00567), (0.00406; 0.00354; 0.00327)]   &       $(2.899; 2.551; 2.394)\times 10^{-17}$  \\
20  & $3\%$    & [(-0.00874; -0.00832; -0.00815), (0.00613; 0.00586; 0.00574)]   &       $(4.076; 3.871; 3.801)\times 10^{-17}$  \\
20  & $5\%$    & [(-0.01043; -0.01004; -0.00987), (0.00781; 0.00757; 0.00745)]   &       $(4.941; 4.831; 4.770)\times 10^{-17}$  \\
\hline
50  & $0\%$    & [(-0.00564; -0.00509; -0.00483), (0.00304; 0.00264; 0.00243)]   &       $(2.307; 2.030; 1.905)\times 10^{-17}$  \\
50  & $3\%$    & [(-0.00845; -0.00812; -0.00798), (0.00584; 0.00566; 0.00597)]   &       $(3.911; 3.758; 3.710)\times 10^{-17}$  \\
50  & $5\%$    & [(-0.01028; -0.00994; -0.00979), (0.00766; 0.00747; 0.00737)]   &       $(4.941; 4.775; 4.726)\times 10^{-17}$  \\
\hline
100  & $0\%$   & [(-0.00502; -0.00454; -0.00432), (0.00242; 0.00209; 0.00192)]    &     $(1.940; 1.707; 1.602)\times 10^{-17}$  \\
100  & $3\%$   & [(-0.00834; -0.00805; -0.00793), (0.00573; 0.00559; 0.00552)]    &     $(3.852; 3.718; 3.678)\times 10^{-17}$  \\
100  & $5\%$   & [(-0.01023; -0.00991; -0.00976), (0.00761; 0.00744; 0.00735)]    &     $(4.912; 4.756; 4.710)\times 10^{-17}$  \\
\hline
110  & $0\%$  & [(-0.00494; -0.00447; -0.00426), (0.00234; 0.00202; 0.00186)]     &     $(1.895; 1.667; 1.564)\times 10^{-17}$  \\
110  & $3\%$  & [(-0.00833; -0.00804; -0.00792), (0.00572; 0.00559; 0.00551)]     &     $(3.846; 3.714; 3.675)\times 10^{-17}$  \\
110  & $5\%$  & [(-0.01022; -0.00990; -0.00976), (0.00760; 0.00744; 0.00735)]     &     $(4.909; 4.754; 4.709)\times 10^{-17}$  \\
\hline\hline
\end{tabular}
\end{center}
\end{table}

\begin{table}[!ht]
\caption{Model-independent sensitivity estimates for the $\tilde{a}_\tau$ magnetic moment and the $\tilde{d}_\tau$ electric dipole moment
for $Q^2_{max}=2, 16, 64\hspace{0.8mm}GeV^2$, $\sqrt{s}=3\hspace{0.8mm}TeV$ and ${\cal L}=50, 100, 200, 300, 450\hspace{0.8mm}fb^{-1}$ at $95\%$ C.L.,
through the process $\mu^{+}\mu^{-} \rightarrow \mu^{+}\gamma^{*} \gamma^{*} \mu^{-} \rightarrow \mu^{+}\tau \bar{\tau}\mu^{-}$.}
\begin{center}
 \begin{tabular}{ccccc}
\hline\hline
\multicolumn{4}{c}{ $\sqrt{s}=3\hspace{0.8mm}TeV$, \hspace{5mm}  $95\%$ C.L.}\\
 \hline
 \cline{1-4}  ${\cal L}\hspace{0.8mm}(fb^{-1})$  &  $\delta_{sys}$ &     $\tilde{a}_\tau$   &  $|\tilde{d}_\tau(e cm)|$ \\
\hline
50  &  $0\%$   & [(-0.00473; -0.00430; -0.00408), (0.00256; 0.00224; 0.00207)]   &       $(1.928; 1.727; 1.620)\times 10^{-17}$   \\
50  &  $3\%$   & [(-0.00756; -0.00728; -0.00714), (0.00540; 0.00523; 0.00515)]   &       $(3.543; 3.434; 3.373)\times 10^{-17}$   \\
50  &  $5\%$   & [(-0.00926; -0.00893; -0.00878), (0.00710; 0.00690; 0.00680)]   &       $(4.495; 4.369; 4.299)\times 10^{-17}$   \\
\hline
100   & $0\%$     & [(-0.00421; -0.00383; -0.00365), (0.00203; 0.00177; 0.00164)]    &      $(1.622; 1.453; 1.363)\times 10^{-17}$  \\
100   & $3\%$     & [(-0.00750; -0.00723; -0.00710), (0.00533; 0.00518; 0.00511)]    &      $(3.505; 3.407; 3.351)\times 10^{-17}$  \\
100   & $5\%$     & [(-0.00921; -0.00891; -0.00876), (0.00703; 0.00688; 0.00678)]    &      $(4.476; 4.356; 4.288)\times 10^{-17}$  \\
\hline
200   & $0\%$     & [(-0.00378; -0.00345; -0.00329), (0.00160; 0.00139; 0.00128)]    &      $(1.365; 1.222; 1.146)\times 10^{-17}$  \\
200   & $3\%$     & [(-0.00746; -0.00720; -0.00708), (0.00530; 0.00516; 0.00509)]    &      $(3.485; 3.393; 3.340)\times 10^{-17}$  \\
200   & $5\%$     & [(-0.00921; -0.00890; -0.00875), (0.00705; 0.00687; 0.00677)]    &      $(4.467; 4.350; 4.283)\times 10^{-17}$  \\
\hline
300   &  $0\%$    & [(-0.00356; -0.00326; -0.00312), (0.00139; 0.00120; 0.00111)]    &      $(1.233; 1.051; 1.035)\times 10^{-17}$  \\
300   &  $3\%$    & [(-0.00745; -0.00720; -0.00707), (0.00528; 0.00515; 0.00508)]    &      $(3.478; 3.389; 3.337)\times 10^{-17}$  \\
300   &  $5\%$    & [(-0.00920; -0.00890; -0.00875), (0.00704; 0.00686; 0.00677)]    &      $(4.464; 4.350; 4.281)\times 10^{-17}$  \\
\hline
450   &  $0\%$    &  [(-0.00337; -0.00310; -0.00296), (0.00120; 0.00103; 0.00095)]   &     $(1.114; 0.998; 0.936)\times 10^{-17}$  \\
450   &  $3\%$    &  [(-0.00744; -0.00719; -0.00707), (0.00528; 0.00515; 0.00508)]   &     $(3.474; 3.385; 3.334)\times 10^{-17}$  \\
450   &  $5\%$    &  [(-0.00920; -0.00889; -0.00875), (0.00704; 0.00686; 0.00677)]   &     $(4.462; 4.347; 4.280)\times 10^{-17}$  \\
\hline\hline
\end{tabular}
\end{center}
\end{table}

\begin{table}[!ht]
\caption{Model-independent sensitivity estimates for the $\tilde{a}_\tau$ magnetic moment and the $\tilde{d}_\tau$ electric dipole moment
for $Q^2_{max}=2, 16, 64\hspace{0.8mm}GeV^2$, $\sqrt{s}=6\hspace{0.8mm}TeV$ and ${\cal L}=50, 100, 300, 500, 710\hspace{0.8mm}fb^{-1}$ at $95\%$ C.L.,
through the process $\mu^{+}\mu^{-} \rightarrow \mu^{+}\gamma^{*} \gamma^{*} \mu^{-} \rightarrow \mu^{+}\tau \bar{\tau}\mu^{-}$.}
\begin{center}
 \begin{tabular}{ccccc}
\hline\hline
\multicolumn{4}{c}{ $\sqrt{s}=6\hspace{0.8mm}TeV$, \hspace{5mm}  $95\%$ C.L.}\\
 \hline
 \cline{1-4}  ${\cal L}\hspace{0.8mm}(fb^{-1})$  &  $\delta_{sys}$ &     $\tilde{a}_\tau$   &  $|\tilde{d}_\tau(e cm)|$ \\
\hline
50  &  $0\%$   & [(-0.00410; -0.00371; -0.00353), (0.00220; 0.00196; 0.00303)]    &      $(1.533; 1.505; 1.415)\times 10^{-17}$   \\
50  &  $3\%$   & [(-0.00686; -0.00657; -0.00643), (0.00501; 0.00487; 0.00479)]    &      $(3.263; 3.153; 3.093)\times 10^{-17}$   \\
50  &  $5\%$   & [(-0.00839; -0.00815; -0.00789), (0.00658; 0.00640; 0.00630)]    &      $(4.134; 4.000; 3.926)\times 10^{-17}$   \\
\hline
100  & $0\%$   & [(-0.00365; -0.00332; -0.00315), (0.00174; 0.00155; 0.00145)]    &      $(1.407; 1.266; 1.191)\times 10^{-17}$  \\
100  & $3\%$   & [(-0.00682; -0.00654; -0.00640), (0.00496; 0.00484; 0.00476)]    &      $(3.236; 3.134; 3.077)\times 10^{-17}$  \\
100  & $5\%$   & [(-0.00837; -0.00804; -0.00788), (0.00655; 0.00638; 0.00628)]    &      $(4.122; 3.991; 3.917)\times 10^{-17}$  \\
\hline
300  &  $0\%$  & [(-0.00310; -0.00282; -0.00269), (0.00119; 0.00105; 0.00098)]    &      $(1.070; 0.963; 0.906)\times 10^{-17}$  \\
300  &  $3\%$  & [(-0.00678; -0.00651; -0.00639), (0.00493; 0.00481; 0.00474)]    &      $(3.218; 3.121; 3.067)\times 10^{-17}$  \\
300  &  $5\%$  & [(-0.00836; -0.00803; -0.00787), (0.00654; 0.00637; 0.00627)]    &      $(4.113; 3.985; 3.914)\times 10^{-17}$  \\
\hline
500   & $0\%$    & [(-0.00290; -0.00264; -0.00252), (0.00098; 0.00087; 0.00081)]    &     $(0.942; 0.848; 0.798)\times 10^{-17}$  \\
500   & $3\%$    & [(-0.00678; -0.00615; -0.00638), (0.00492; 0.00481; 0.00474)]    &     $(3.215; 3.118; 3.064)\times 10^{-17}$  \\
500   & $5\%$    & [(-0.00835; -0.00803; -0.00787), (0.00654; 0.00637; 0.00627)]    &     $(4.112; 3.984; 3.913)\times 10^{-17}$  \\
\hline
710   & $0\%$    & [(-0.00278; -0.00253; -0.00242), (0.00086; 0.00076; 0.00071)]    &     $(0.863; 0.777; 0.731)\times 10^{-17}$  \\
710   & $3\%$    & [(-0.00677; -0.00651; -0.00639), (0.00492; 0.00481; 0.00474)]    &     $(3.213; 3.117; 3.063)\times 10^{-17}$  \\
710   & $5\%$    & [(-0.00835; -0.00803; -0.00787), (0.00653; 0.00637; 0.00627)]    &     $(4.111; 3.983; 3.913)\times 10^{-17}$  \\
\hline\hline
\end{tabular}
\end{center}
\end{table}

\section{Conclusions}

\vspace{3mm}

We perform a comprehensive study of the sensitivity to both the total cross-section of the process
$\mu^{+}\mu^{-} \rightarrow \mu^{+}\gamma^{*} \gamma^{*} \mu^{-} \rightarrow \mu^{+}\tau \bar{\tau}\mu^{-}$
and on the $\tilde{a}_\tau$ magnetic moment and the $\tilde{d}_\tau$ electric dipole moment, with respect to the parameters of the
future muon collider, $\sqrt{s}$ and ${\cal L}$, as well as of the virtuality of the photon $Q^2_{max}$. We consider the most general
Lagrangian coupling of two tau-leptons and the photon (see Eqs. (7)-(13)), which involves both $\tilde{a}_\tau$ and $\tilde{d}_\tau$
interactions.

We found that, for ${\tilde a}_\tau$ and ${\tilde d}_\tau$, the best sensitivity constraints come from consider $\sqrt{s}=6\hspace{0.8mm}TeV$,
${\cal L}= 710\hspace{0.8mm}fb^{-1}$ and $Q^2_{max}= 64\hspace{0.8mm}GeV^2$ and we estimated the sensitivity to be
${\tilde a}_\tau= (-0.00242, 0.00071)$ and $|{\tilde d}_\tau|= 0.731\times 10^{-17}$  at $95\%$ C.L. as is show in Table III.
This compares favorably with earlier DELPHI and BELLE studies for $\tau$MM and $\tau$EDM (see Eqs. (1) and (2)), and readily
provides leading sensitivity for $\tilde{a}_\tau$ and $\tilde{d}_\tau$.

We already show through Figs. 2-19 and Tables I-III that a future $\mu^+\mu^-$ collider, currently envisioned as a machine for
new physics BSM, will have leading sensitivity to probing both $\tilde{a}_\tau$ and $\tilde{d}_\tau$ couplings simultaneously
through the process $\mu^{+}\mu^{-} \rightarrow \mu^{+}\gamma^{*} \gamma^{*} \mu^{-} \rightarrow \mu^{+}\tau \bar{\tau}\mu^{-}$.
However, it is worth mentioning that significant room remains to be explored in both the $\tilde{a}_\tau$ and $\tilde{d}_\tau$
couplings.

In conclusion, our study complement and extend previous $\tilde{a}_\tau$ and $\tilde{d}_\tau$ sensitivity estimates made for
various specific collider environments. In general, the muon collider with the highest integrated luminosity in proposal can
reach better sensitivity couplings in the high energy region. In addition, this collider with larger center-of-mas energies is able to
explore broader parameter space in the high energy region.

\vspace{8mm}

\begin{center}
{\bf Acknowledgements}
\end{center}

M. K. and A. A. B. acknowledge that this work is supported by the Scientific Research Project Found of Cumhuriyet University under
the project number TEKNO-022. A. G. R. and M. A. H. R. acknowledge support from SNI and PROFOCIE (M\'exico).

\vspace{2cm}


\newpage

\begin{figure}[H]
\centerline{\scalebox{0.6}{\includegraphics{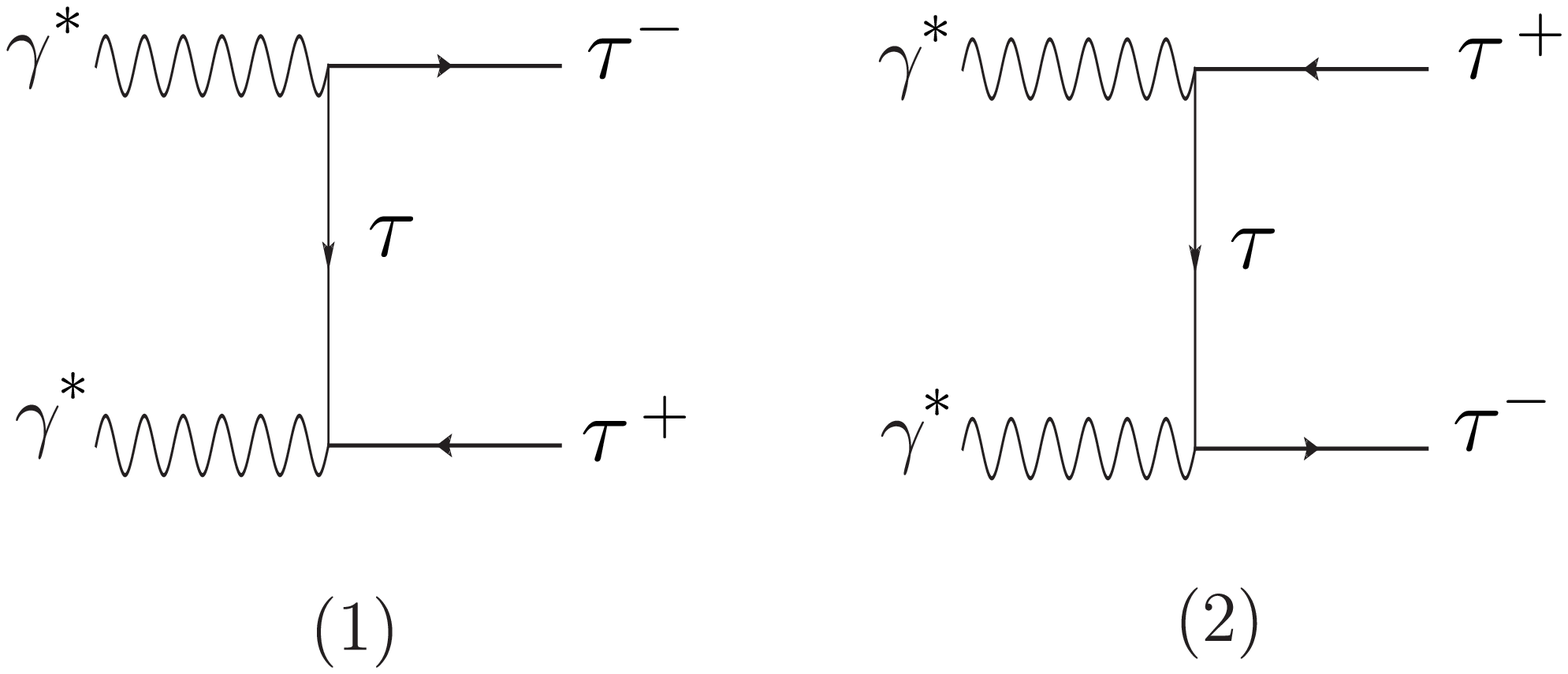}}}
\caption{ \label{fig:gamma} The Feynman diagrams contributing to the subprocess
$\gamma^* \gamma^* \to \tau^+ \tau^-$.}
\end{figure}

\begin{figure}[H]
\centerline{\scalebox{1.2}{\includegraphics{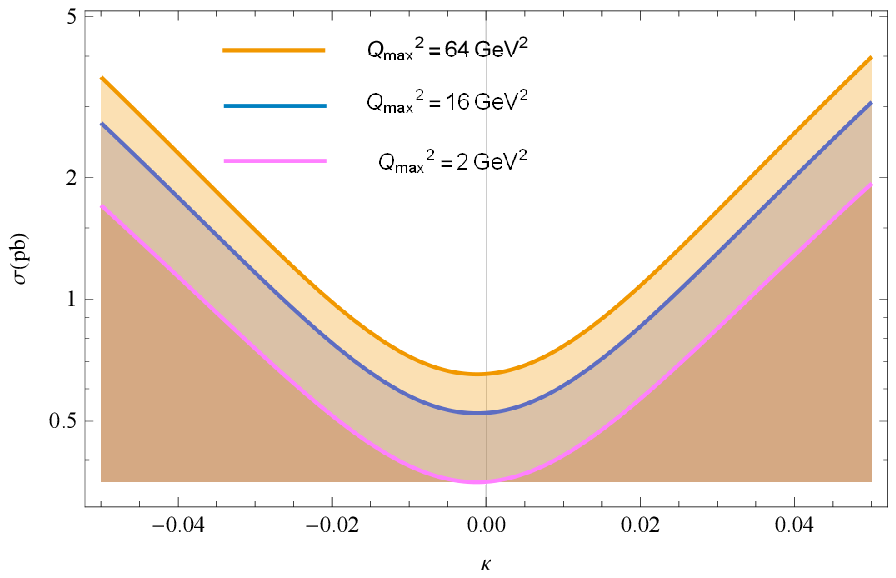}}}
\caption{ \label{fig:gamma1} The total cross-sections of the process
$\mu^{+}\mu^{-} \rightarrow \mu^{+}\gamma^{*} \gamma^{*} \mu^{-} \rightarrow \mu^{+}\tau \bar{\tau}\mu^{-}$
as a function of $\kappa$ for center-of-mass energy of $\sqrt{s}=1.5\hspace{0.8mm}TeV$.}
\end{figure}

\begin{figure}[H]
\centerline{\scalebox{1.2}{\includegraphics{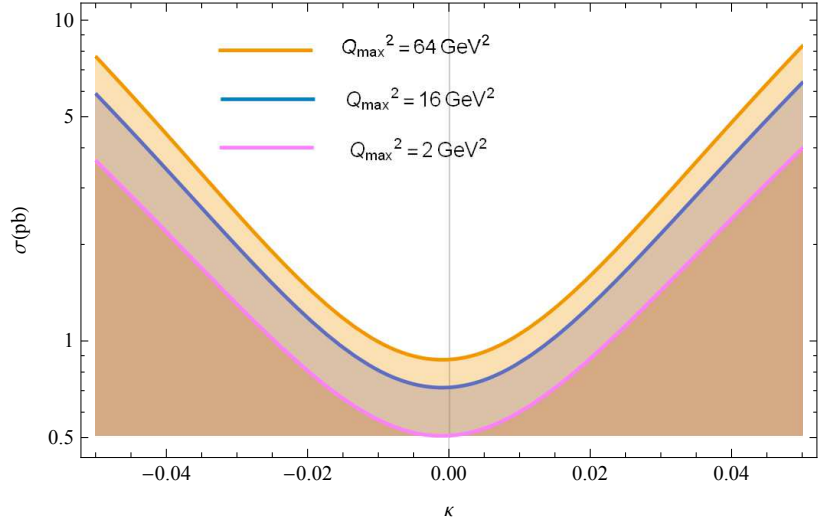}}}
\caption{ \label{fig:gamma2} Same as in Fig. 2, but for
$\sqrt{s}=3\hspace{0.8mm}TeV$.}
\end{figure}

\begin{figure}[H]
\centerline{\scalebox{1.2}{\includegraphics{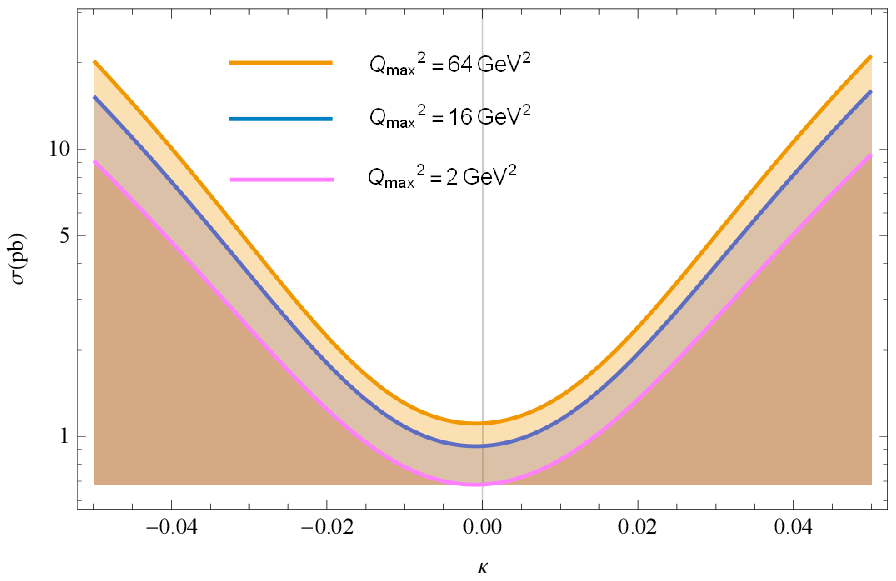}}}
\caption{ \label{fig:gamma3} The same as Fig. 2, but for
$\sqrt{s}=6\hspace{0.8mm}TeV$.}
\end{figure}

\begin{figure}[H]
\centerline{\scalebox{1.2}{\includegraphics{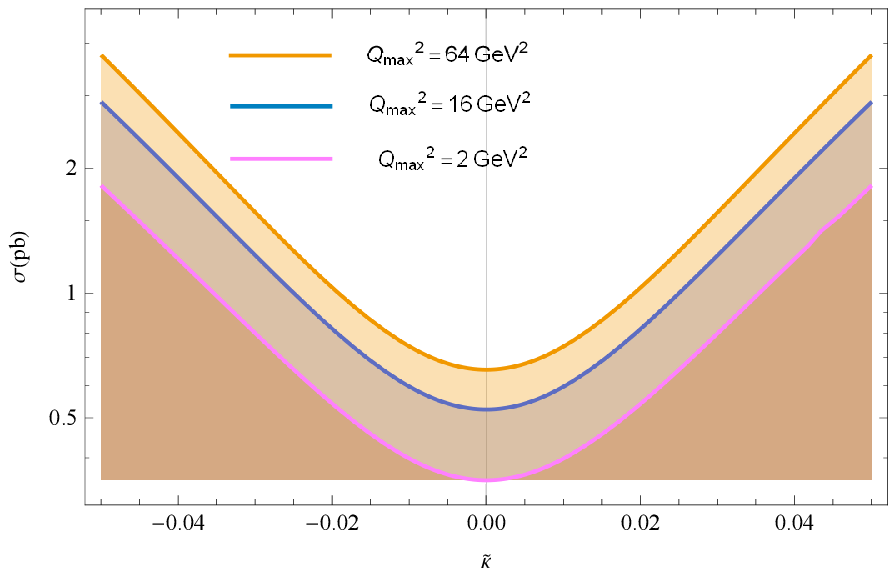}}}
\caption{ \label{fig:gamma4}
The total cross-sections of the process
$\mu^{+}\mu^{-} \rightarrow \mu^{+}\gamma^{*} \gamma^{*} \mu^{-} \rightarrow \mu^{+}\tau \bar{\tau}\mu^{-}$
as a function of $\tilde\kappa$ for center-of-mass energy of $\sqrt{s}=1.5\hspace{0.8mm}TeV$.}
\end{figure}

\begin{figure}[H]
\centerline{\scalebox{1.2}{\includegraphics{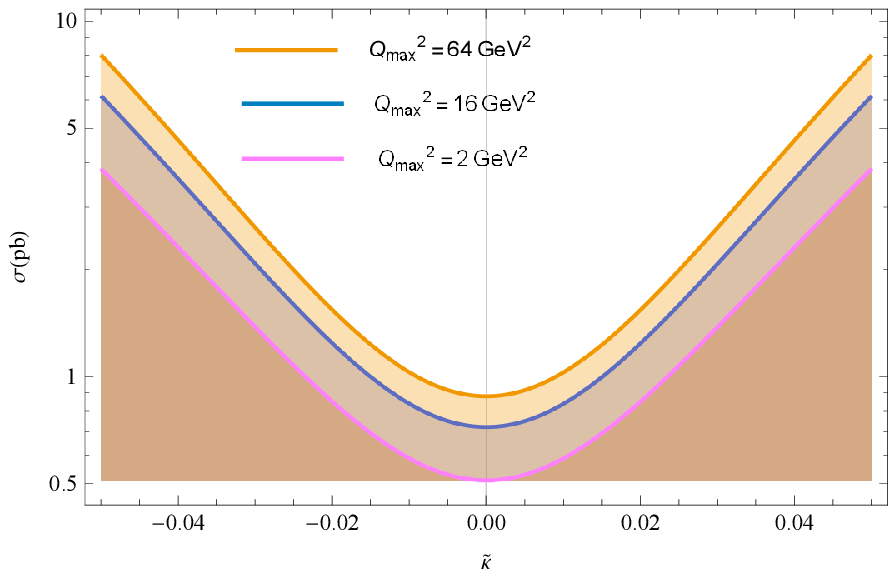}}}
\caption{ \label{fig:gamma15} The same as Fig. 5, but for
$\sqrt{s}=3\hspace{0.8mm}TeV$.}
\end{figure}

\begin{figure}[H]
\centerline{\scalebox{1.2}{\includegraphics{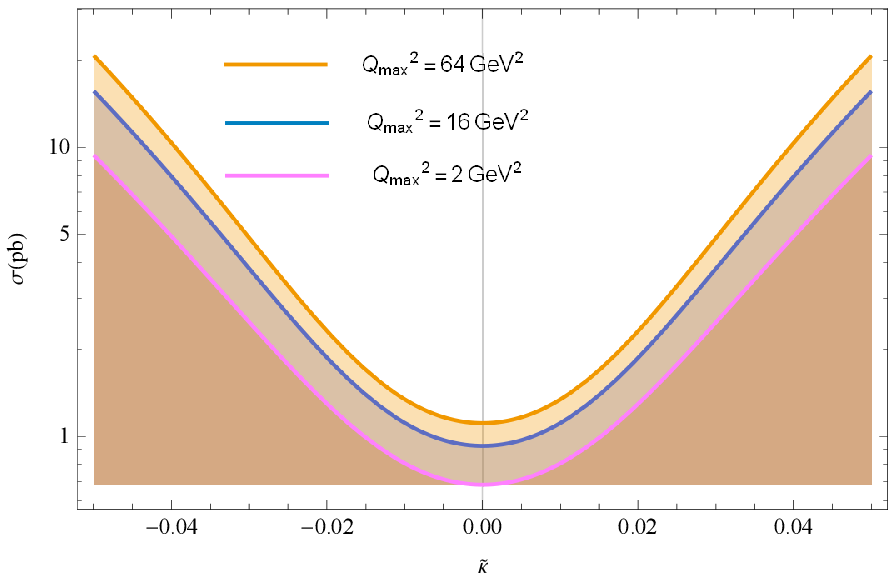}}}
\caption{\label{fig:gamma16} The same as Fig. 5, but for
$\sqrt{s}=6\hspace{0.8mm}TeV$.}
\end{figure}

\begin{figure}[H]
\centerline{\scalebox{1.2}{\includegraphics{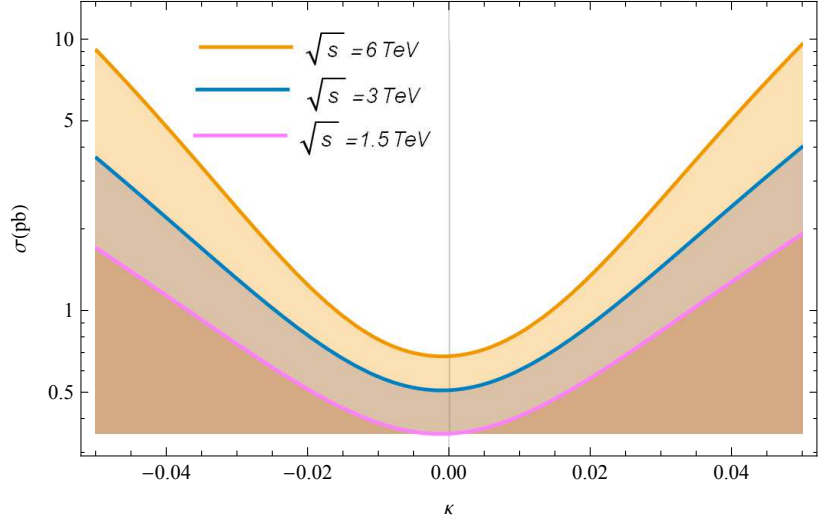}}}
\caption{\label{fig:gamma17} The total cross-sections of the process
$\mu^{+}\mu^{-} \rightarrow \mu^{+}\gamma^{*} \gamma^{*} \mu^{-} \rightarrow \mu^{+}\tau \bar{\tau}\mu^{-}$
as a function of $\kappa$ for $Q^2_{max}=2\hspace{0.8mm}GeV^2$.}
\end{figure}

\begin{figure}[H]
\centerline{\scalebox{1.2}{\includegraphics{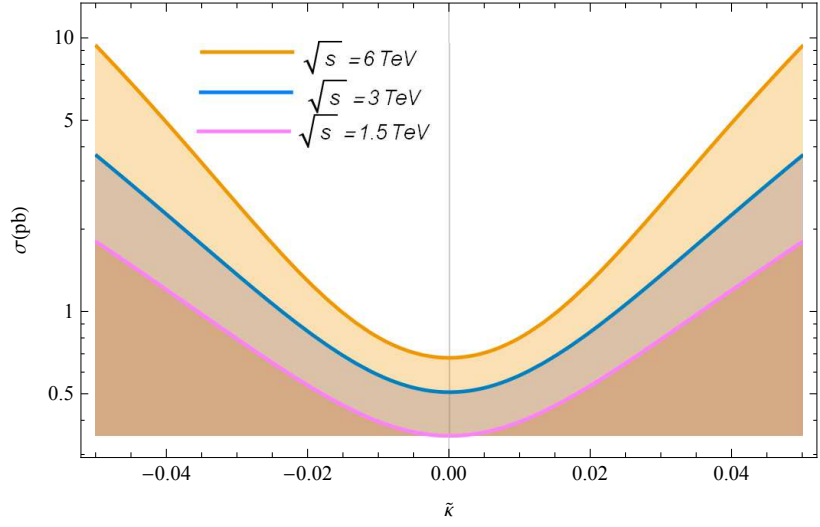}}}
\caption{ \label{fig:gamma5} The same as Fig. 8, but for
$\tilde\kappa$.}
\end{figure}

\begin{figure}[H]
\centerline{\scalebox{1.2}{\includegraphics{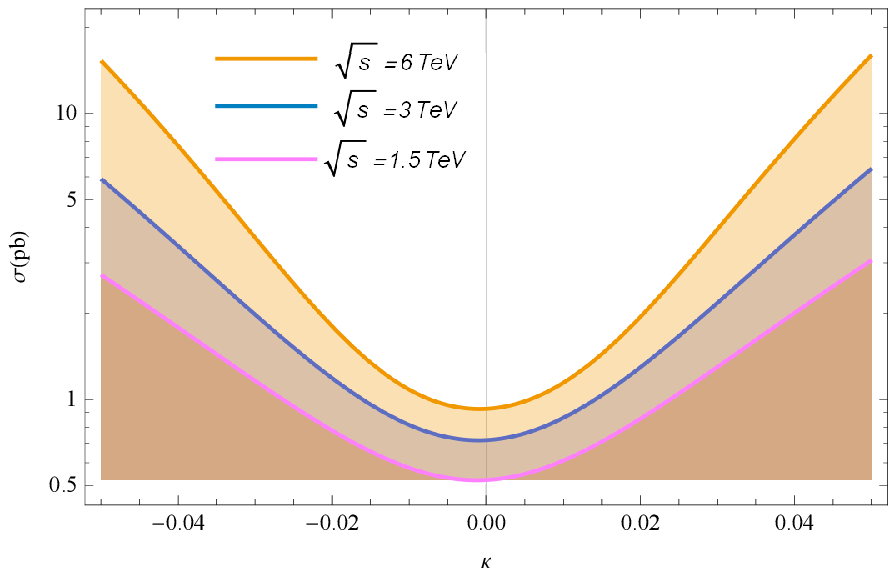}}}
\caption{\label{fig:gamma17} The same as Fig. 8, but for $Q^2_{max}=16\hspace{0.8mm}GeV$.}
\end{figure}

\begin{figure}[H]
\centerline{\scalebox{1.2}{\includegraphics{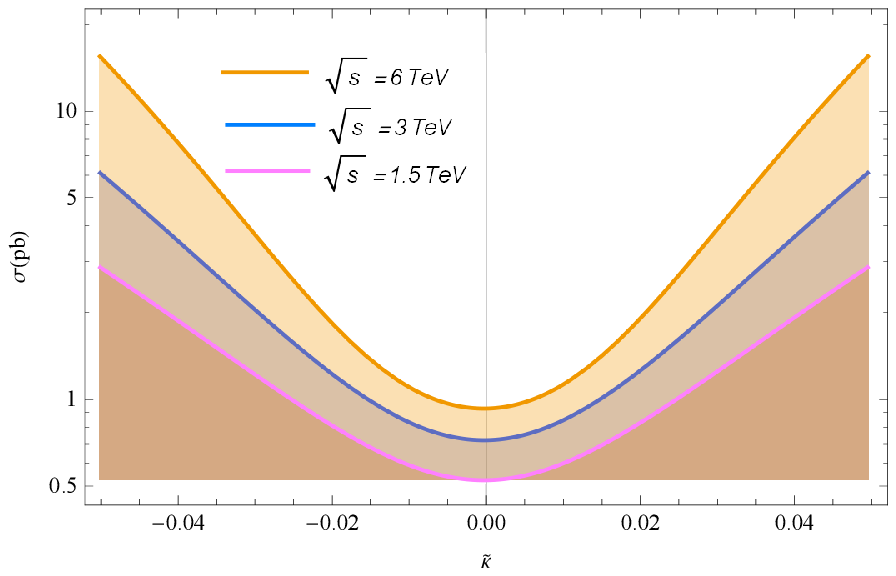}}}
\caption{\label{fig:gamma17} The same as Fig. 9, but for $Q^2_{max}=16\hspace{0.8mm}GeV$.}
\end{figure}

\begin{figure}[H]
\centerline{\scalebox{1.2}{\includegraphics{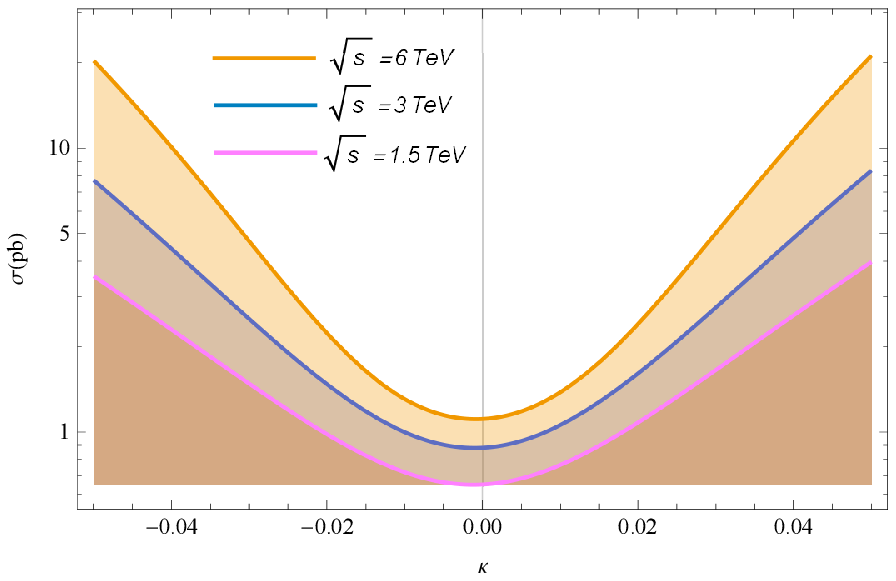}}}
\caption{\label{fig:gamma17} The same as Fig. 8, but for $Q^2_{max}=64\hspace{0.8mm}GeV$.}
\end{figure}

\begin{figure}[H]
\centerline{\scalebox{1.2}{\includegraphics{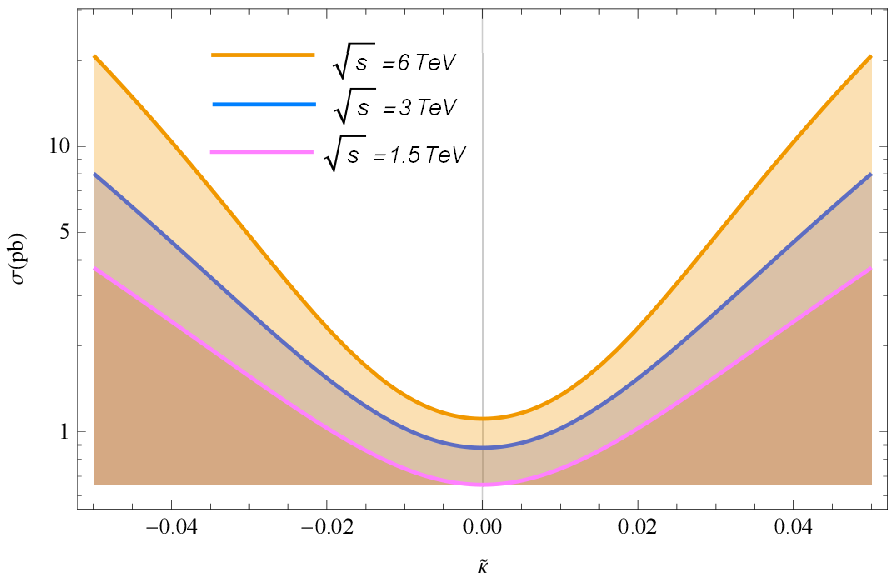}}}
\caption{\label{fig:gamma17} The same as Fig. 9, but for $Q^2_{max}=64\hspace{0.8mm}GeV$.}
\end{figure}

\begin{figure}[H]
\centerline{\scalebox{0.9}{\includegraphics{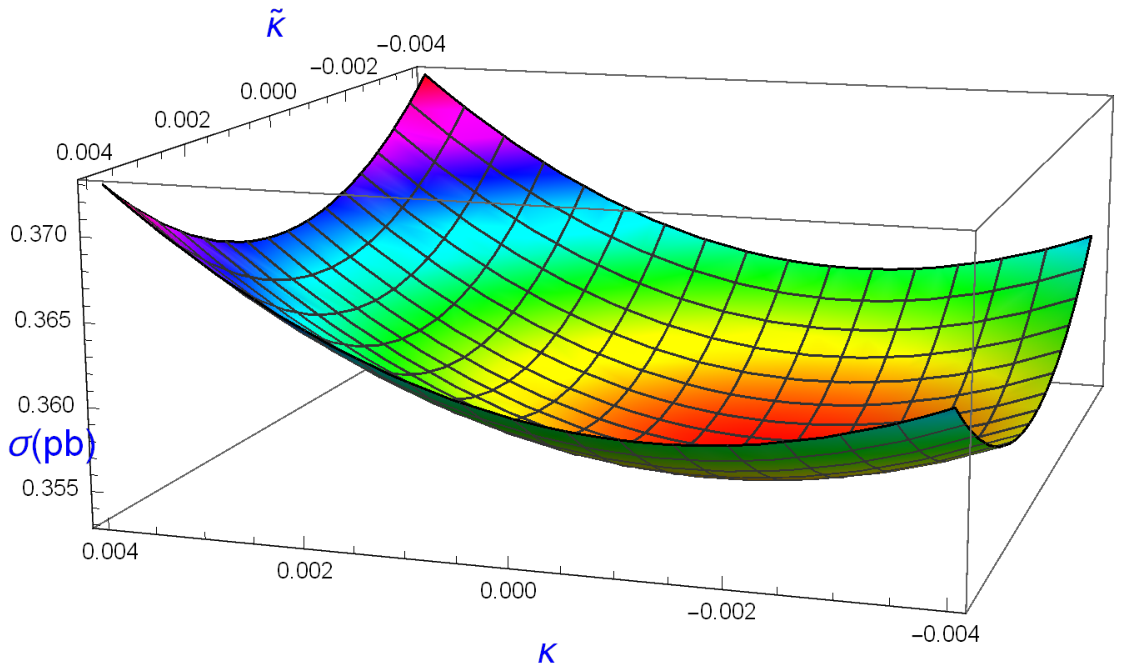}}}
\caption{\label{fig:gamma17} The total cross-sections of the process
$\mu^{+}\mu^{-} \rightarrow \mu^{+}\gamma^{*} \gamma^{*} \mu^{-} \rightarrow \mu^{+}\tau \bar{\tau}\mu^{-}$
as a function of $\kappa$ and $\tilde\kappa$ for center-of-mass energy $\sqrt{s}=1.5\hspace{0.8mm}TeV$ and
$Q^2=2\hspace{0.8mm}GeV^2$.}
\end{figure}

\begin{figure}[H]
\centerline{\scalebox{0.9}{\includegraphics{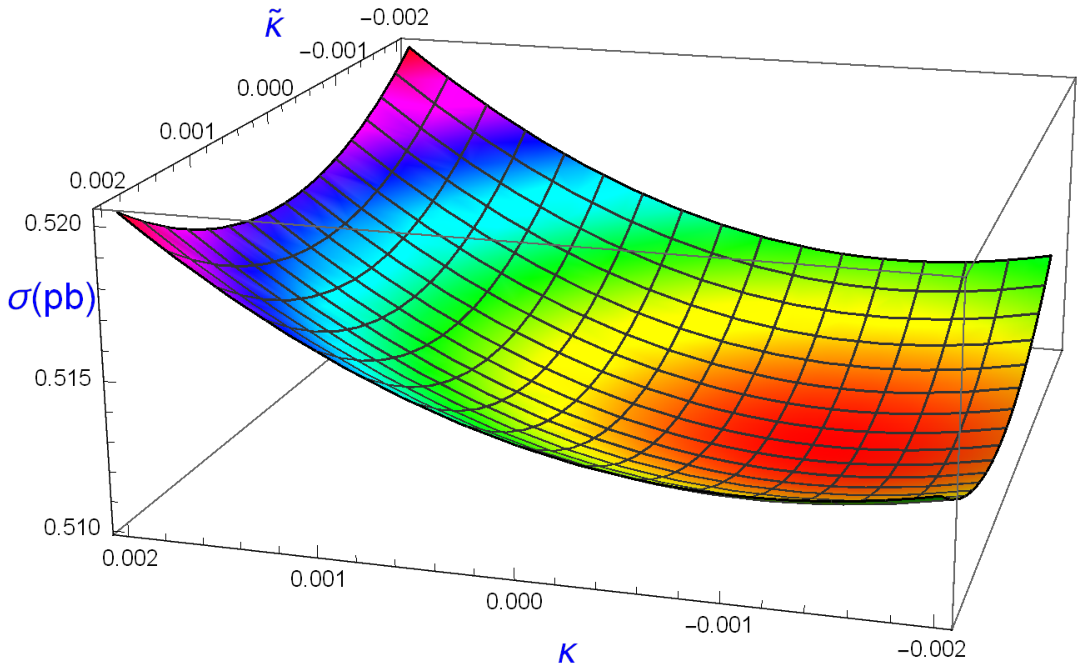}}}
\caption{\label{fig:gamma17} The same as Fig. 14, but for $\sqrt{s}=3\hspace{0.8mm}TeV$.}
\end{figure}

\begin{figure}[H]
\centerline{\scalebox{0.9}{\includegraphics{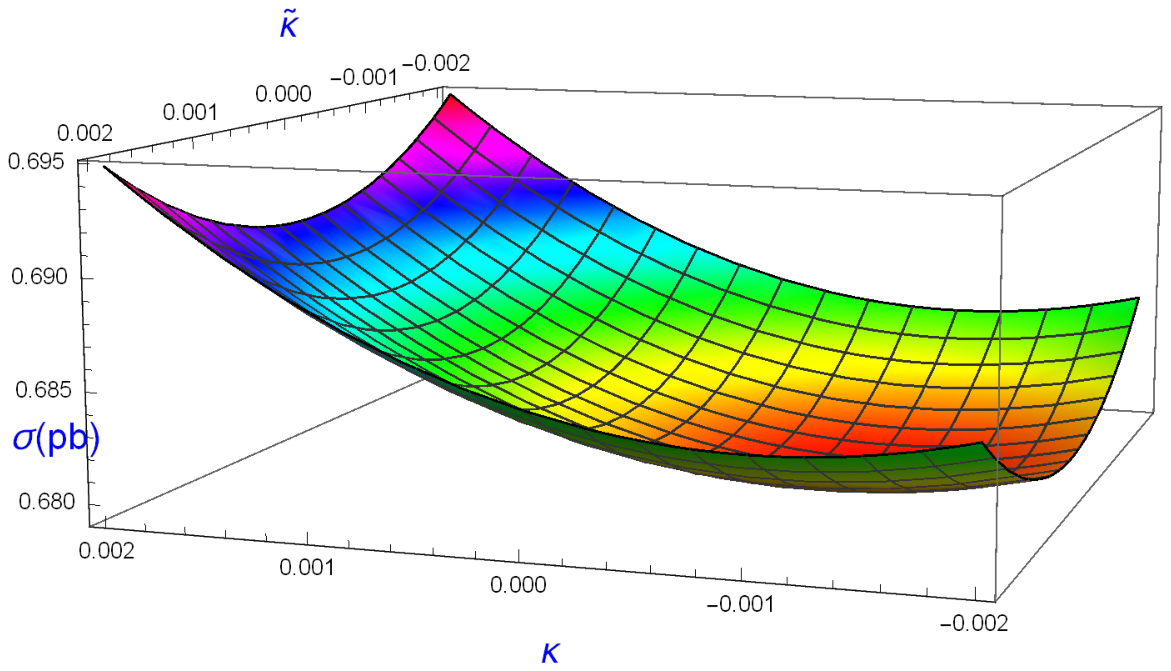}}}
\caption{\label{fig:gamma17} The same as Fig. 14, but for $\sqrt{s}=6\hspace{0.8mm}TeV$.}
\end{figure}

\begin{figure}[H]
\centerline{\scalebox{1}{\includegraphics{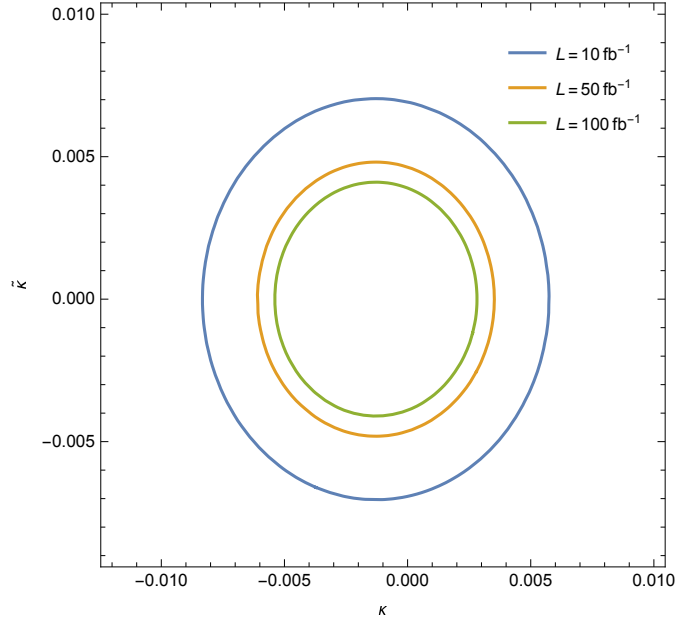}}}
\caption{ \label{fig:gamma6} Sensitivity contours at the $95\%\hspace{0.8mm}C.L.$ in the
$(\kappa-\tilde\kappa)$ plane for the process
$\mu^{+}\mu^{-} \rightarrow \mu^{+}\gamma^{*} \gamma^{*} \mu^{-} \rightarrow \mu^{+}\tau \bar{\tau}\mu^{-}$
for center-of-mass energy $\sqrt{s}=1.5\hspace{0.8mm}TeV$ and $Q^2=2\hspace{0.8mm}GeV^2$.}
\end{figure}

\begin{figure}[H]
\centerline{\scalebox{1}{\includegraphics{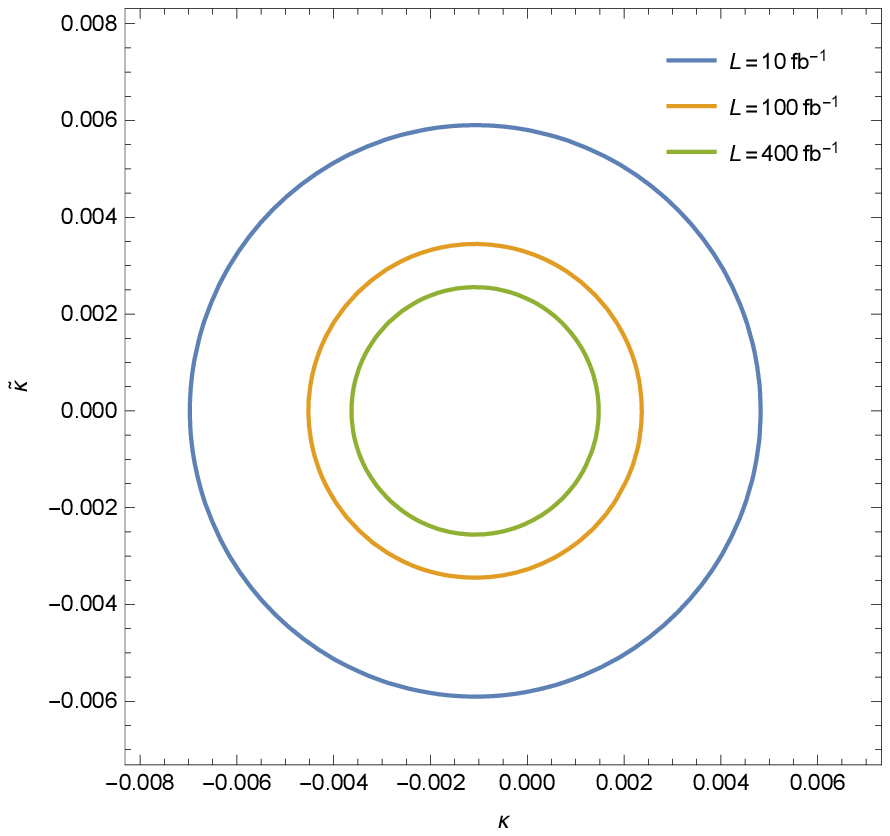}}}
\caption{\label{fig:gamma18} The same as Fig. 17, but for
$\sqrt{s}=3\hspace{0.8mm}TeV$.}
\end{figure}

\begin{figure}[H]
\centerline{\scalebox{1}{\includegraphics{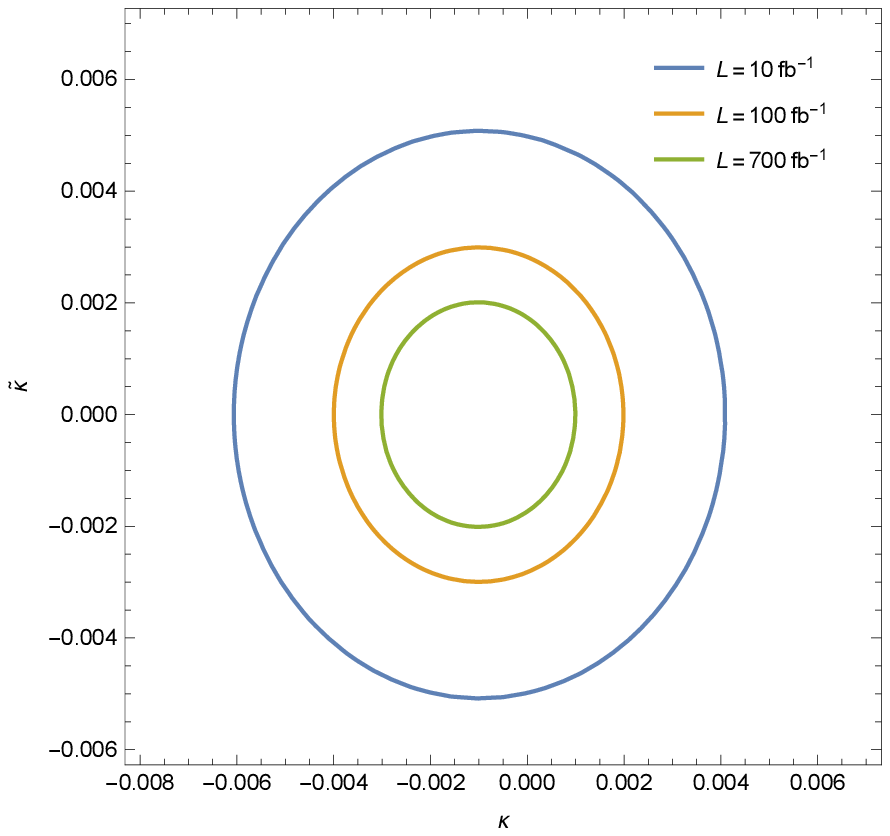}}}
\caption{\label{fig:gamma17} The same as Fig. 17, but for
$\sqrt{s}=6\hspace{0.8mm}TeV$.}
\end{figure}

\end{document}